\documentclass[twocolumn,superscriptaddress,prb,aps,amsmath,amssymb,showpacs,lengthcheck]{revtex4-1} 

\usepackage{graphicx}

\begin{document}
\DeclareGraphicsExtensions{.eps}

\title{Microscopic Model for Intersubband Gain from Electrically Pumped Quantum-Dot Structures}

\author{Stephan Michael}
\affiliation{Department of Physics and Research Center OPTIMAS, University of Kaiserslautern, P.O. Box 3049, 67653 Kaiserslautern, Germany}
\author{Weng W. Chow}
\affiliation{Semiconductor Materials and Device Sciences Department, Sandia National Laboratories, Albuquerque, NM 87185-1086, USA} 
\author{Hans Christian Schneider}
\affiliation{Department of Physics and Research Center OPTIMAS, University of Kaiserslautern, P.O. Box 3049, 67653 Kaiserslautern, Germany}

\begin{abstract}
We study theoretically the performance of electrically pumped self-organized quantum dots as a gain material in the mid-IR range at room temperature. We analyze an AlGaAs/InGaAs based structure composed of dots-in-a-well sandwiched between two quantum wells.  
We numerically analyze a comprehensive model by combining a many-particle approach for electronic dynamics with a realistic modeling of the electronic states in the whole structure. We investigate the gain both for quasi-equilibrium conditions and current injection. Comparing different structures we find that steady-state gain can only be realized by an efficient extraction process, which prevents an accumulation of electrons in continuum states, that make the available scattering pathways through the quantum dot active region too fast to sustain inversion. The tradeoff between different extraction/injection pathways is discussed. 
Comparing the modal gain to a standard quantum-well structure as used in quantum cascade lasers, our calculations predict reduced threshold
current densities of the quantum dot structure for comparable modal gain. Such a comparable modal gain can, however, only be achieved for an inhomogeneous broadening of a quantum-dot ensemble that is close to the lower limit achievable today using self-organized growth.
\end{abstract}

\maketitle

\section{Introduction\label{Introduction}}

Much research on light emitters in the mid-infrared range has been focused on
quantum cascade lasers (QCL),~\cite{QCL1,QCL2,QCL3,QCL4,QCL5,Harrison1,Harrison2,Harrison6,Ines1,Ines2}
which are complex structures consisting of hundreds
of coupled quantum wells (QWs). QCLs can produce a high output power
and operate up to and above room temperature.~\cite{MIR4,MIR3,MIR2,MIR5,QCL6,QCL7}
QWs usually emit light only in-plane due to the transverse magnetic (TM)
polarization of the intersubband transition. 
To achieve emission perpendicular to the surface from intersubband transitions one needs to fabricate
wavelength specific surface output couplers.~\cite{MIR6} 
Mid-infrared lasers emit in a frequency range close to thermal
energies, so that there may be considerable thermal energy losses. The development of more efficient emitters is therefore an important problem.~\cite{QCL3,Harrison6}  The use of
nanostructures with a three dimensional confinement leads to discrete level
energies and thus limits the phase space for the interaction with phonons,
which makes nonradiative recombinations much less likely.~\cite{bottleneck1,bottleneck2,MIR13}  For instance, a magnetic field  leads
to an increased efficiency of QCLs due to the occurrence of quantized electronic Landau levels.~\cite{MIR8,Harrison8}  Also a quantum-dash cascade structure was proposed.~\cite{quantum-dash} 

Another possibility is to use self-assembled quantum dots (QDs).\cite{QDcascade} Experimental
results have demonstrated nonradiative relaxation times that are orders of
magnitude longer than in QW structures.~\cite{intro11,MIR12,intro13}
There have been studies of midinfrared photodetectors using QDs,~\cite{MIR15,MIR16} and optimization issues have been addressed.~\cite{MIR2:3} In addition the room temperature ultraweak absorption of a single
buried semiconductor QD was measured.~\cite{MIR2:5} Also type-II
InAsSb/InAs QDs for midinfrared applications have been investigated ~\cite{MIR2:1} and midinfrared photoluminescence of epitaxial PbTe/CdTe
QDs has been studied.~\cite{MIR2:4} 

QD intersubband transitions are particularly promising for mid-infrared wavelengths.~\cite{intraband-DWELL-PDetector}
These transitions allow light emission normal to the growth
direction. Additionally, they are a basic requirement for the realization of a
QCL consisting of QDs. Steps in this direction include the demonstration of midinfrared
electroluminescence at low temperatures,~\cite{MIR2:6,MIR19,MIR20} and a theoretical proposal of TE-polarized optical gain through a ruby-type three level scheme.~\cite{rubyscheme} While QD midinfrared
emission of devices using additional interband transitions has also been
proposed,~\cite{MIR18} such a scheme would be only feasible for weak cavity fields.

Recently, progress in room temperature midinfrared electroluminescence from QDs was made.~\cite{midinfrared1,el_roomtemp}
An essential part of the proposed structure in Ref.~\onlinecite{midinfrared1}   
is electron tunneling between QW and QD states. The properties of related tunneling processes between localized and continuum states in self-organized QD structures have been separately investigated.~\cite{tunneling_th,tunneling_exp} 

The present paper presents a theoretical model for a structure similar to
Ref.~\onlinecite{midinfrared1}. We assume a heterostructure consisting of a
thin active layer of QDs embedded in a QW (a so-called DWELL
structure), which is, in turn sandwiched between two QWs. We refer to this as
a QW-QD-QW heterostructure. 
Because QW-QD-QW heterostructures include a dots-in-a-well (DWELL) structure,
not only electron tunneling between QW and QD states 
but also typical effects of density-dependent carrier dynamics for DWELL
heterostructures are of importance.~\cite{prasankumar}
Using electronic eigenstates for the whole structure as input, we 
solve the dynamical equations for the electronic level occupations and for the
important coherences in the system under investigation.
In doing so we distinguish between intra-QD, intra-QW and QD-QW electron
scattering and calculate the underlying electron-phonon and electron-electron 
scattering processes microscopically following a many-particle approach that includes, in particular, the effects of the electron-phonon interactions on the QD states~\cite{PQE_Review}. 
Using different models for the excitation process, we determine the achievable
inversion (gain) in the active medium. 
Based on the numerical results, we discuss possible optimizations of the
design of AlGaAs/InGaAs QW-QD-QW structures as active material for midinfrared lasers.
To the best of our knowledge, a microscopic theoretical investigation and optimization 
for the capability of those devices for midinfrared laser applications is still missing.

In the present paper we focus
on some physical properties and parameter dependencies associated with the gain achievable in QDs by current injection. This does not solve all the numerous technical challenges of a QD laser device but may contribute some design criteria for future devices. Furthermore, we present a comparison between QD-QCL and QW-QCL devices. 
We  focus on the active material and do not include collector regions as in QD-QCLs. However, the results of the analysis done are
transferable to periodic structures (such as a QD-QCL), if small carrier losses in the collector region are neglected. 


This paper is organized as follows. In Sec.~\ref{gainmaterial1} we describe
two QW-QD-QW structures that represent possible designs for a gain material in
the infrared, and calculate the electronic band lineup and
wave functions. There is a brief review of the semiconductor Bloch equations
and their scattering contributions that we use to describe the structures
under consideration in Sec.~\ref{SBE}. In Sec.~\ref{numerical1} we investigate
the conditions for which inversion between the ground and degenerate excited
states in the QDs can be achieved, assuming fixed carrier densities in the
QWs. The small signal gain is determined from the population inversion. In
Sec.~\ref{numerical2} we incorporate carrier injection (extraction) into the
model and compare the different structures. In Sec.~\ref{numerical3},
we present numerical results for stronger fields and identify a range of
parameters for which gain in the midinfrared at room temperature is
feasible. Additionally, the tradeoff between the different injection (extraction) pathways 
and the consequences of leakage are discussed in Sec.~\ref{numerical2add}. Finally, in Sec.~\ref{numerical2add2}, we compare a standard QW-QCL design from Ref.~\onlinecite{Harrison1} to our QD-QCL alternative to illustrate the possible potential of QD-QCLs.

\section{Electronic structure of a QW-QD-QW system\label{gainmaterial1}}

\begin{figure}[tb]
\centering
\includegraphics[trim=5cm 3cm 4.5cm 3cm,clip,scale=0.5]{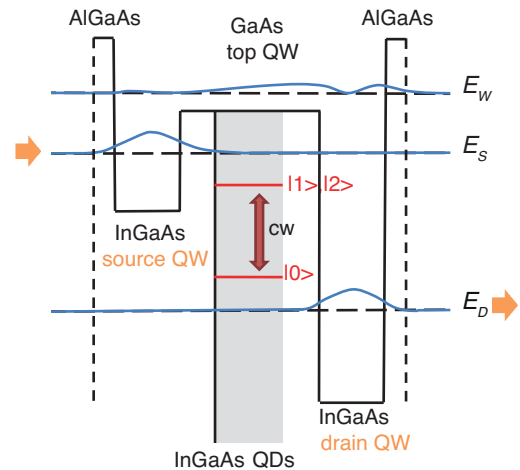}
\caption{Confinement potential in growth direction (black), QW wave functions (blue) and optically active states (red) for structure A, which is optimized for wave-function overlap, see text. Carrier injection and extraction processes are indicated by arrows (orange).}
\label{skizze_3well}
\end{figure}

\begin{figure}[tb]
\centering
\includegraphics[trim=5cm 3cm 4.5cm 3cm,clip,scale=0.5]{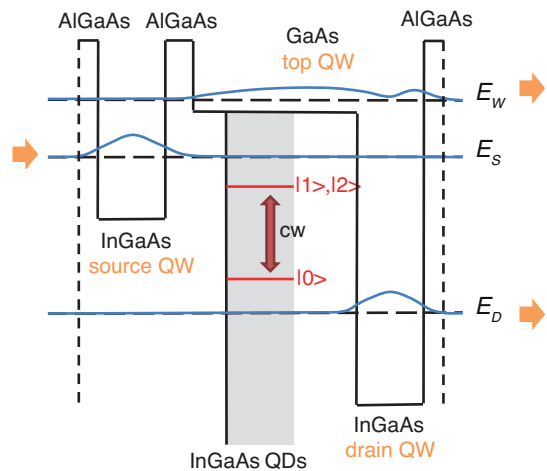}
\caption{Confinement potential in growth direction (black), QW wave functions
  (blue) and optically active states (red) for structure B, which includes a
  barrier between source QW and top QW. Carrier injection and extraction
  processes are indicated by arrows (orange). Note the additional carrier extraction from the top QW states compared to Fig.~\ref{skizze_3well}}.
\label{skizze_3well2}
\end{figure}

We describe here a QW-QD-QW system with a QD layer (DWELL structure) designed
to emit in the midinfrared and potentially suitable as intersubband-laser gain 
medium. Fig.~\ref{skizze_3well} shows a structure designed to rely on electron-phonon scattering for creating
inversion in the QDs. 

For the structure ``A'' in Fig.~\ref{skizze_3well} we assume a cw field resonant with the transition between the lowest electronic level $| 0\rangle $ of the QD and the
excited level $| 1\rangle $, $|2\rangle$, which are degenerate. 
If a quasi-equilibrium Fermi-Dirac 
distribution is maintained in the DWELL structure, a population inversion for the optically active states in the QD is not
possible in steady-state, so that it is necessary to extract carriers out of
the lowest electronic level of the DWELL structure. 
This is achieved by an additional ``drain'' QW with electronic band
edge $E_{D}$  that is offset roughly by a LO phonon energy from the lowest electronic level $E_0$ in the QD, i.e., $E_{0}-E_{D}=\hbar \omega_{\text{LO}}+\epsilon $, where $\epsilon>0$ is
smaller than a few meV. We keep $\epsilon$ in the calculation because a perfect
lineup of the structure is not necessary. Since the wave-functions of the
QD and drain QW states do not overlap appreciably, the scattering
process is slow compared to a similar scattering process between the
extended states and localized states in the DWELL structure. We refer to the extended states in the DWELL structure as ``top'' QW even though they are not pure plane waves, but have been orthogonalized to the localized QD levels. In particular, electron-electron
scattering between QD and top QW states for a significant occupation of
the top QW is extremely efficient. If the source for carriers is the
top QW, the relaxation from the QD state $| 0\rangle $
to the states of the drain QW will be not efficient enough to extract electrons from level $| 0\rangle $, and thus keep the transition $|0\rangle \leftrightarrow |1\rangle$ inverted. Carrier injection is therefore done in our structure from a second QW, referred to as ``source'' QW with a electronic band
edge $E_{S}$ offset by an LO phonon energy from the excited levels $|1\rangle
$ and $|2\rangle$, i.e., $E_{S}-E_{1,2}=\hbar \omega _{LO}-\delta $, where $\delta>0$
is also smaller than a few meV. We further assume in the following an energy
difference $E_{1,2}-E_{0}\gg \hbar\omega_{\text{LO}}$, which leads to a so-called phonon bottleneck effect because
transitions between the discrete electron states $|0\rangle \leftrightarrow|1\rangle,|2\rangle$ are inhibited.~\cite{bottleneck1,MIR13} To facilitate steady-state
population inversion for the optical active states electron-electron scattering processes that are assisted by transitions in the top QW should be suppressed as much as possible. This is achieved by the energy difference $E_{W}-E_{1,2}\gg \hbar \omega _{LO}$ between the excited QD levels and the band edge $E_W$ of 
the top QW. With such a band lineup the
carrier-density of the top QW and the electron-electron scattering contribution from this density is kept as small as possible. 

With the band lineup described so far, it remains to optimize the injection and removal of
carriers for the operation as a light emitter. To this end, the source and the drain QW
wave functions need to have significant overlap with the QD wave functions,
but the layers cannot be too close to each other to avoid electrical breakdown between the
source- and drain-QW. 

As a variation of the structure ``design'' we will also consider carrier removal from the extended states in the top QW in addition to the removal process through the QD states. Removal of carriers is provided through subbands of the surrounding heterostructure. In the structures analyzed here, the composition and shape of the electronic structure leads to a top QW with an admixture of the first excited subband of the drain QW, which realizes an efficient overlap of the drain QW \emph{and top QW} with the surrounding heterostructure.  A small width of the source- and drain-QW helps to increase the overlap further. However, in our investigation we do not include the design of the surrounding heterostructure, which is indicated by the broken lines at the left and right side of the band lineups in Figs.~\ref{skizze_3well} and~\ref{skizze_3well2}. 

We now present in some details the geometry and material parameters used for the calculation of the electronic structures shown in Fig.~\ref{skizze_3well}, which incorporates the design principles discussed so far. We assume an ensemble of In$_{0.75}$Ga$_{0.25}$As QDs on a wetting
layer with a thickness of $1$ nm embedded in the GaAs top QW. The
geometry of the QDs is a truncated pyramid with $\left\{ 101\right\} $
facets. The QDs have a base of $12\times 12$~nm and height of $3$~nm. For
an overlap-optimized structure (see Fig.~\ref{skizze_3well}) the GaAs
top QW has a width of $10$~nm and the bottom of the wetting layer has a
distance of $3$ nm to the source QW. The In$_{0.12}$Ga$_{0.88}$As
source QW and the In$_{0.38}$Ga$_{0.62}$As drain QW have both a width of 
$5$~nm. The whole system is embedded in an Al$_{0.1}$Ga$_{0.9}$As barrier.

The electronic structure is calculated by $k\cdot p$ theory~\cite{homepage} as
described in Appendix \ref{a_mparameter}.  For computational reasons, we treat the calculation of the three-dimensional QD states separately
from the calculation of the one-dimensional envelope of the QWs, and orthogonalize the
three-dimensional QW states to describe the whole system. For the QD we obtain a
ground and two degenerate excited states. For the source-, the top- and the drain-QW
only one confined subband exists, respectively. The excited drain-QW subband
is mixed with the top QW confined subband as discussed above. For the
combined system the line up of states are compiled in Table~\ref{tabel-1}.
The transition energy between the optical active states of the QD
is $E_{1,2}-E_{0}=105$~meV. This corresponds to a mid-infrared wavelength of $11.8\,\mu$m.

\begin{table}[bt]
\begin{tabular}{|l|l|c|}
\hline
Band edge of \ldots & Symbol & Energy (meV) \\ \hline
top QW & \multicolumn{1}{|c|}{$E_{W}$} & $+12$ \\ \hline
source QW & \multicolumn{1}{|c|}{$E_{S}$} & $-55$ \\ \hline
drain QW & \multicolumn{1}{|c|}{$E_{D}$} & $-230$ \\ \hline
\hline
QD state & Symbol & Energy (meV) \\ \hline
$\left\vert 1\right\rangle $, $\left\vert 2\right\rangle $ & 
\multicolumn{1}{|c|}{$E_{1,2}$} & $-85$ \\ \hline
$\left\vert 0\right\rangle $ & \multicolumn{1}{|c|}{$E_{0}$} & $-190$ \\ 
\hline
\end{tabular}
\caption{Compilation of the line up of states for the combined QW-QD-QW system measured
  against the bottom of the top QW potential.}
\label{tabel-1}
\end{table}

\begin{figure}[tb]
\centering
\includegraphics[trim=7cm 5.8cm 7cm 5.2cm,clip,scale=0.5]{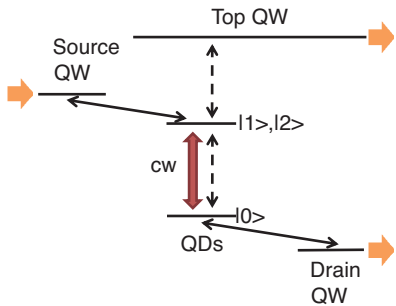}
\caption{Schematic picture of the carrier dynamics in the QW-QD-QW system.
Indicated are: (i) injection/extraction process (horizontal orange arrows),
(ii) transitions dipole-coupled to the optical field (red vertical arrow),  
(iii) scattering processes in the DWELL structure (dashed thin arrows,) and
(iv) scattering processes between source/drain QWs and the QD (thin arrows).}
\label{skizze_QWQDQW}
\end{figure}

For comparison, structure ``B'', shown in Fig.~\ref{skizze_3well2} is introduced, which is less aggressively optimized for wave-function overlap and incorporates some safeguards against electrical breakdown and current leakage. To this end the distance between source- and
drain-QW is increased, and an Al$_{0.1}$Ga$_{0.9}$As barrier between the
source QW and the top QW is introduced. In addition, the barrier 
between source QW and top QW allows both QWs to be addressed 
separately by an injection and extraction processes. In particular, in structure~B carriers can be extracted from the drain QW and the
top QW, as indicated in Figs.~\ref{skizze_3well2} and~\ref{skizze_QWQDQW}.
The barrier has a width of $2$~nm and the total distance between source and
drain QW is 14\,nm. The wetting layer is 5\,nm above the source QW. To
obtain comparable energies, we corrected the composition of the source QW
to In$_{0.13}$Ga$_{0.87}$As for the numerical calculation. All other
parameters remain unchanged, including the carrier injection process. Note, however, that for structure A we assume carrier extraction from the drain QW only. In the following, we thus compare the performance of structures with optimized wave-function overlap (structure A) and optimized carrier extraction (structure B).

\section{Semiconductor Bloch equations \label{SBE}}

The dynamics of the polarizations and carrier distributions at the
single-particle level are calculated in the framework of the
semiconductor Bloch equations for the reduced single-particle density
matrix. We denote in the following electron levels in the QD with $\alpha$.
For the optical active states of interest one obtains the following
equations of motion for the ``intra(electron-)band'' polarizations~$p_{\alpha\alpha'}$
\begin{eqnarray}
\frac{\partial }{\partial t}p_{\alpha\alpha'}
&=& - \left( i\omega _{\alpha'\alpha} + \gamma_{d} \right) p_{\alpha\alpha'} 
-i\Omega_{\alpha'\alpha}
\left( n_{\alpha'}-n_{\alpha}\right) 
\label{p-aa}
\end{eqnarray}
where $\gamma_{d}$ is a decay rate for the polarization.
For the time evolution of the electron populations
$n_{\alpha }$ one obtains
\begin{eqnarray}
\frac{\partial}{\partial t}n_{\alpha} &=&
i\sum\limits_{\alpha ^{\prime }\neq \alpha }\left( \Omega
_{\alpha \alpha'}\text{ }p_{\alpha \alpha'
}-\Omega _{\alpha' \alpha}\text{ }p_{\alpha' \alpha}\right)
+S_{\alpha}
\label{n-a} 
\end{eqnarray}
The coherent contributions of the above equations containing the transition
frequencies $\omega_{\alpha' \alpha }$ and Rabi
frequencies $\Omega_{\alpha \alpha'} = \hbar^{-1} \mu_{\alpha \alpha'} E
\left( t \right)  $ where $ E \left( t \right) $ is the electric field at the
position of the QD.

The term $S_{\alpha}$ describes the scattering contributions in the
dynamical equations for the electron distributions and contains the
influence of electron-electron Coulomb $S_{\alpha}^{\text{cc}}$ and electron-phonon
$S_{\alpha}^{\text{cp}}$ scattering. Their theoretical treatment is
contained in the following section.

In the semiconductor Bloch equations~\eqref{p-aa}  and \eqref{n-a} also Hartree-Fock energy renormalizations arise,
which can reach a few meV for highly populated QD states. However, energy
shifts of only a few meV do not affect the scattering behavior significantly.
Moreover, the Hartree-Fock energy renormalization has the same effect on the
steady-state result of the population inversion as a slight change of the
material composition. An optimization of the electronic structure including Hartree-Fock energy
renormalizations would require inverse quantum-engineering as described in Ref.~\onlinecite{Ines1}, which is beyond the scope of the present paper. We therefore neglect renormalization due to Coulomb interaction. For the calculation with an optical field in Sec.~\ref{numerical3}, we are mainly interested in the qualitative dependence on the optical field intensity, which is treated as a parameter in our calculation. Thus we also neglect Hartree-Fock contributions that result in and of the Rabi energy, which would have to be included in a more comprehensive calculation where the dynamics of the optical field is also included.
 
\subsection{Scattering contributions}

The scattering contribution $S_{\alpha}$ includes both electron-electron and
electron-phonon scattering.  Our treatment is described in more detail in Appendix \ref{a_scatter}, where the explicit expressions are given. Here we only summarize our approach. 

While electrons interact with longitudinal acoustic (LA) and
longitudinal optical (LO) phonons, 
scattering effects due to acoustic phonons in QDs  are estimated to be
very inefficient,\cite{SQD} as long as level spacing of the QDs is
much larger as the typical energy range of the acoustic phonons coupled to the
QDs, i.e., below a few meV in InGaAs QDs or QD molecules.\cite{Giannozi,Zimmermann1}

Scattering processes involving QD states connect discrete levels so that the
influence of level broadening is much more pronounced than for scattering 
between continuum states in QWs. Thus, we follow
Ref.~\onlinecite{QDM} and introduce 
an \emph{effective quasi-particle broadening} for the scattering contributions.  
By using an effective quasi-particle broadening we work with polarons, i.e., quasi-particles that include the effect of the 
coupling to phonons, instead of the ``naked'' QD electronic levels.
We have determined this broadening from single-pole approximations to the zero-density QD
polaronic spectral functions, see also in Ref.~\onlinecite{QDM}, in the style of Ref.~\onlinecite{Jahnke4,Jahnke2} 
and neglected the Coulomb-interaction contribution to the effective quasi-particle broadening. This is a
valid approximation, if the continuum states, i.e., especially the top QW, are not
appreciably populated by carriers, which is necessary if gain, i.e., inversion, for the optically
active transition is desired, see Sec.~\ref{numerical1}.
 
A constant level broadening around $\Gamma = 0.5\,\text{meV}$, i.e., $\Gamma \approx \hbar \times 0.75 \,\text{ps}^{-1}$, was calculated for typical InAs QDs.~\cite{dissertation}
Here, we assume the level broadening of a typical InAs QD, 
because a precise calculation of the level broadening of 
the QD in our QW-QD-QW structure is too demanding 
with respect to computing time.
The QD under investigation has rather a large level spacing. That is why the stated
value for the broadening tends to result to an overestimation.
Because a small broadening reduces the electron-phonon relaxation from the excited to
the ground state of the QD, the gain is reduced by an overestimation of the
broadening. Thus, to be on the safe side its better to slightly overestimate rather than to underestimate the
broadening of the QD states.
All in all, the precise value of $\Gamma$ does not affect the statements of
our theoretical analysis, but it is important to get its order of magnitude right. 

With the considerations above it turns out that the relaxation or scattering for the carrier distributions cannot easily be computed using Fermi's Golden
Rule arguments because there is no straightforward energy conservation for
transitions between polarons.
Thus, the calculated constant level broadening $\Gamma$ referring to the effect of the electron-phonon interaction 
on the polaronic spectrum in the form of complex renormalized energies of a
single-particle QD state~$\lambda$ has to be incorporated into the 
explicit scattering expressions by
\begin{equation}
\tilde{\epsilon}_{\lambda}=\epsilon_{\lambda}+\Delta\epsilon_{\lambda}-i\Gamma_{\lambda}
\label{complexenergy}
\end{equation}
where $\Delta\epsilon$ is a negligible energy shift (HF correction and a small
correlation contribution). The broadening~$\Gamma_{\lambda}$ of the level $\lambda$ is entirely due to correlations.
This incorporation is done by following Ref. \onlinecite{QDM} for the
derivation of the electron-phonon and electron-electron scattering. 
In contrast to Ref.~\onlinecite{QDM} all coherences
are neglected. This is especially in the case of a small signal gain a valid approximation. We also assume that only conduction band states are involved in
the scattering process, because only electrons in the conduction band are injected and extracted from the
system under investigation. 
The explicit formula expressions for the electron-phonon and electron-electron
scattering is given in Appendix \ref{a_scatter}.

\subsection{Model for carrier injection (extraction)\label{subsec:carrier-injection}}

We next include a simplified model for current injection in the structure described above. We assume that the current injects carriers into the left side of the structure, i.e., the source QW, and removes them from the right side, i.e., drain or top QW.
For an effective injection (extraction) of carriers from a QW, energetically close 
and local nearby subbands have to be provided from the surrounding heterostructure.

For the inclusion of the process, we extend the Bloch equations for the
source QW, according to Ref.~\onlinecite{chowbook1}, by a carrier injection term of the form
\begin{equation} 
\left. \frac{dn_{k}}{dt}\right\vert _{\text{inject}}=\Lambda F_k( 1-n_{k}) 
 \label{midlas_inject}
\end{equation}
where the Pauli blocking factor $(1-n_{k}) $ prevents the pump from injecting
carriers in occupied states. Further, $ \Lambda$ denotes an injection rate and
$F_k$ is a Fermi-Dirac pump distribution.

The pump distribution model in the form~\eqref{midlas_inject} attempts to
capture in a simple form the details of the injection process. It is based on
the assumption that by the time the injected carriers reach the source QW they
have thermalized by collisions and therefore their $k$ dependence can be
described by a quasi-equilibrium pump distribution $F_{k}(N^{F},T^{F})$ with
the characteristic carrier density $N^{F}$ and the characteristic temperature
$T^{F}$ as parameters, which are kept constant. This distribution is weighted by an injection rate $\Lambda$. The temperature $T^{F}$
entering $F_{k}$ is taken to be the lattice temperature. The pump distribution
$F_{k}$ should not be confused with a Fermi-Dirac distribution in the QW.  Note that we use the injection rate $\Lambda$ as the independent parameter and calculate the steady-state current density via $J=e\frac{1}{\mathcal{A}}\sum_k (dn_k/dt)|_{\text{inject}}$ where $\mathcal{A}$ is the normalization area of the quantum well. This expression for the current will be used to compare to similar calculations for quantum well structures in Section~\ref{numerical2add2} below.

To model the extraction of carriers by transport of carriers from the drain or top QW to the right side of the structure, we extend the Bloch equations for the QWs by the simple rate equation
\begin{equation}
\frac{dn_{k}}{dt}\Big\vert_{\text{extract}}=-\Lambda F_{k}n_{k}
 \label{midlas_extract}
\end{equation}
where $n_{k}$ is the occupation of the QW state $k$.

With regard to the injection model we should also briefly discuss changes
introduced to the band lineup in a biased structure. For realistic fields of
several 10\,kV/cm along the growth axis one expects an energy shift of a few meV between nearby QW
and QD states. In agreement with Ref.~\onlinecite{Lim} we neglect these small energy
corrections for the thin QW-QD-QW heterostructure under investigation. For a
potential drop of more than about 15~meV over the active region, the energy
difference between the source QW band bottom and the excited QD level becomes
too large for efficient carrier injection. In this case the design of the
``cold'' structure needs to be changed such that the bias-induced shift leads
to a level lineup close to the one described in Figs.~\ref{skizze_3well} and \ref{skizze_3well2}.
In particular, for a field of 36\,kV/cm as chosen in Sec.~\ref{numerical2add2} 
the ``cold'' structure needs to be changed to
a source-QW composition of In$_{0.19}$Ga$_{0.81}$As and a drain-QW composition
of In$_{0.33}$Ga$_{0.67}$As to obtain approximately the same level lineup as
described above.

\section{Numerical results}

\subsection{Inversion (gain) for fixed QW carrier-densities\label{numerical1}}

In this section we investigate under which conditions regarding the carrier
densities of the QWs an inversion between the ground and degenerate excited
states in the QDs is possible. Therefore we investigate the behavior of the
population inversion in the QDs for fixed quasi-equilibrium distributions in
the QWs. Because the carrier densities in the QWs are kept fixed, no injection
(removal) processes are included.

In the numerical calculation we start with given QW carrier densities and an initially
empty QD system. Importantly, electron-phonon and electron-electron scattering described by
Eqs.~\eqref{electron-phonon} and~\eqref{electron-electron} leads to
QD-QW electron scattering as well as intra-QD scattering processes
(see scattering processes (iii) and (iv) depicted in Fig.~\ref{skizze_QWQDQW}). The carrier distributions are evolved until a steady-state is reached. 

For a weak optical field in resonance with the transition between the lowest and excited states of the QD the steady-state distributions remain unchanged. The intensity gain for such a weak optical fields is given by 
\begin{equation}
G=2\frac{\omega }{c_{0}\varepsilon _{0}n_{b}}\frac{N_{D}}{h_{\text{Rg}}} \frac{\mu ^{2}}{\hbar \gamma _{d}}N  
\label{midlas_gain}
\end{equation}
where $\hbar \omega =105$\,meV is the transition energy, $n_{b}=3.4$ (GaAs)
the background refractive index of the host material, $N_{d}=5\times
10^{10}$~cm$^{-2}$ the QD density, $h_{\text{Rg}}\approx 15$~nm the heights of the
active region, $\mu =2.5e$\,nm the dipole moment, $\hbar \gamma_{d}=1$\,meV
the polarization dephasing and $N$ the inversion of the optically active
states. The intersubband dipole moment $\mu =2.5e$\,nm is five times larger
than the interband dipole moment for the transition between
the electron and hole ground state, which already has an appreciable magnitude. 
Thus,  for the same inversion $N$, the gain on the intersubband transition is larger than that on an interband transition in the QD.
The choice of the polarization dephasing of $\hbar \gamma _{d}=1.0$~meV is motivated by the restrictions
that it has to be higher than intersubband
dephasing for the case of unpopulated QD scattering states and a small carrier density in the QWs
($\hbar \gamma _{d} < 0.1$ meV),\cite{QDM} but lower than an interband
dephasing with an appreciable population in the QD scattering states ($\hbar
\gamma_{d}$ up to $10$ meV).~\cite{highdeph} 
We will plot the small signal gain in addition to the inversion between the optically active states in the following.

\begin{figure}[tb]
\centering
\includegraphics[trim=5cm 2cm 5cm 4cm,clip,scale=0.5]{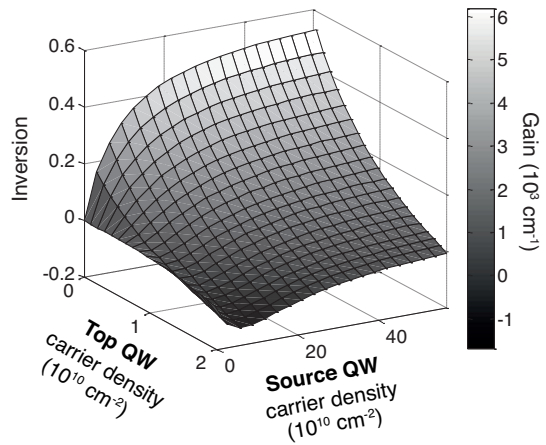}
\caption{Population inversion and gain for the QD transitions in structure A versus
carrier densities in the top and source QW. The lattice temperature is 150\,K.}
\label{num1_fig1}
\end{figure}

\begin{figure}[tb]
\centering
\includegraphics[trim=5cm 2cm 5cm 4cm,clip,scale=0.5]{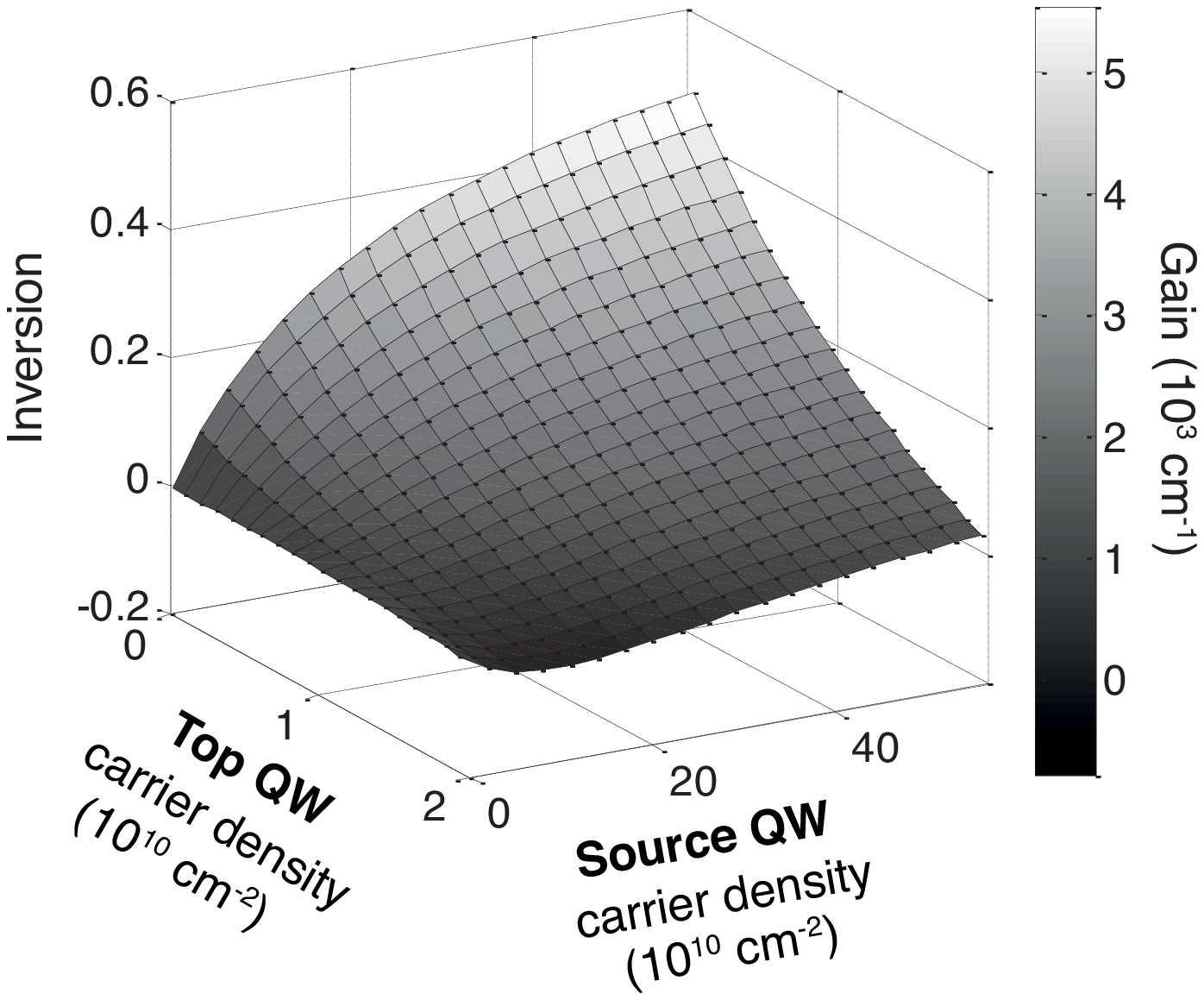}
\caption{Same as Fig.~\ref{num1_fig1} for lattice temperature 300\,K.}
\label{num1_fig2}
\end{figure}

\begin{figure}[tb]
\centering
\includegraphics[trim=5cm 2cm 5cm 4cm,clip,scale=0.5]{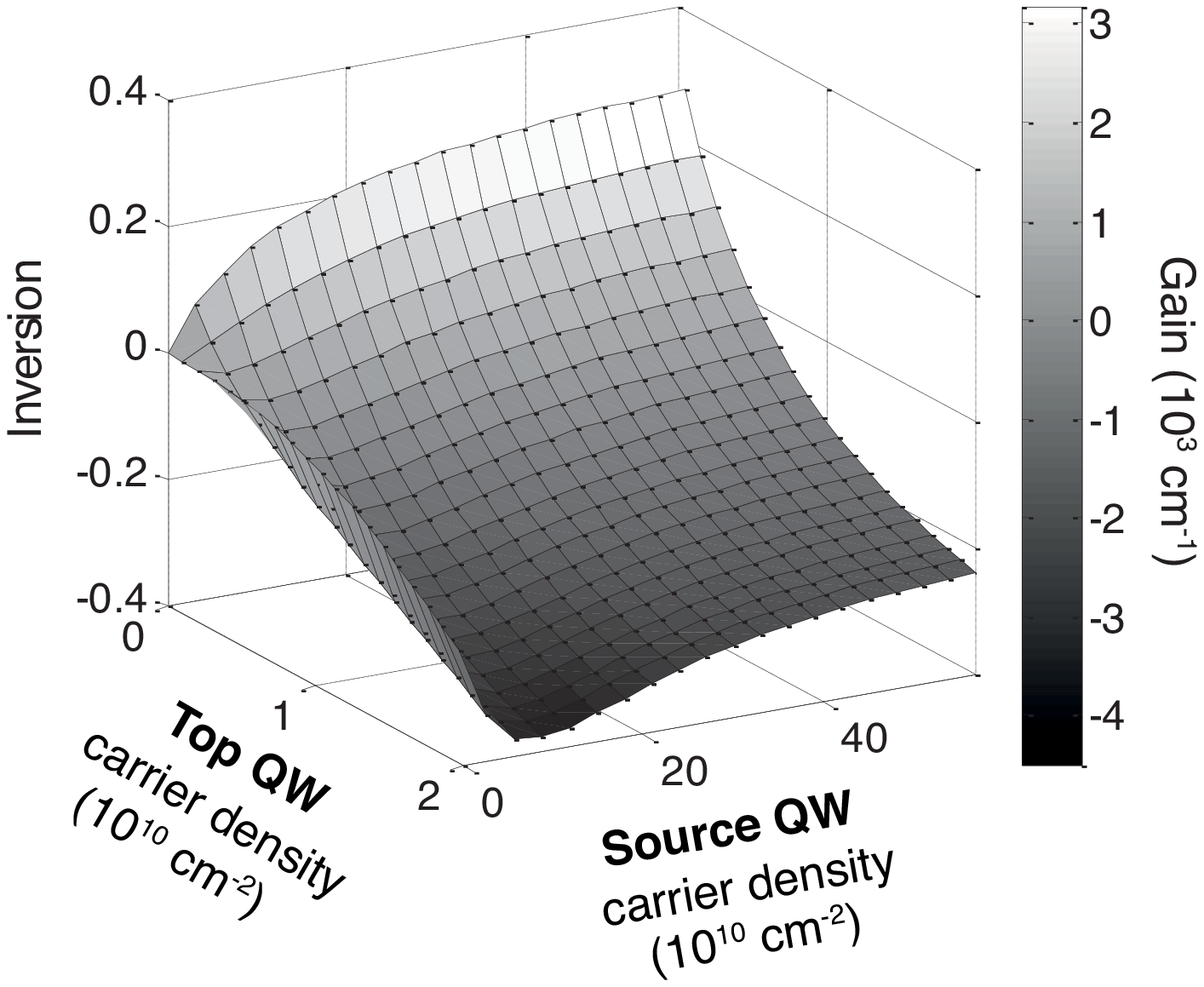}
\caption{Population inversion and gain for the QD transitions in structure B versus
carrier densities in the top and source QW. The lattice temperature is 150~K.}
\label{num1_fig3}
\end{figure}

\begin{figure}[tb]
\centering
\includegraphics[trim=5cm 2cm 5cm 4cm,clip,scale=0.5]{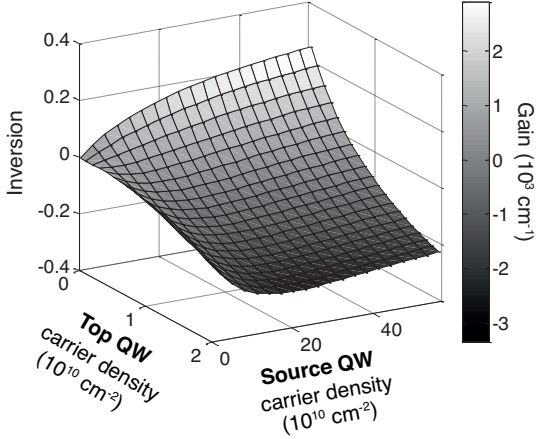}
\caption{Same as Fig.~\ref{num1_fig3} for lattice temperature of 300\,K.}
\label{num1_fig4}
\end{figure}

Figure~\ref{num1_fig1} and Fig.~\ref{num1_fig2} shows the population inversion and gain for the QD
transition in structure A as a function of carrier densities in the top and source QWs. The drain QW is assumed to be empty, which is a ``best case'' assumption for carrier extraction from the active region. In Fig.~\ref{num1_fig1} the lattice temperature is 150\,K. For an empty top QW and a negligible carrier density in the source QW the gain is obviously zero. Up to a source-QW density of $20 \times 10^{10}$~cm$^{-2}$ the gain rises steeply, levels off  in the range between $20 \times 10^{10}$~cm$^{-2}$ to $40 \times 10^{10}$~cm$^{-2}$, and reaches saturation over $40 \times 10^{10}$~cm$^{-2}$.
An increasing carrier density in the top QW for a fixed carrier density in the
source QW leads to a rapid decrease in the gain. For a carrier density of
approximately $2 \times 10^{10}$~cm$^{-2}$ no gain remains, and for higher
densities in the top QW only absorption exists.

Figure~\ref{num1_fig2} depicts the results of a calculation analogous to
Fig.~\ref{num1_fig1}, but for a lattice temperature of 300\,K. The qualitative
analysis remains the same, but the gradient of the gain is lower for
increasing source-QW and top-QW carrier densities. In particular, the gain
reaches saturation for higher carrier densities of the source QW. 

We now repeat these calculations for structure B. Fig.~\ref{num1_fig3} and
Fig.~\ref{num1_fig4} show the results for lattice temperatures of 150\,K and
300\,K, respectively. The overall dependence of the gain on the source-QW and
top-QW densities for structure B is similar to that of structure A. However,
the saturated gain is clearly smaller and the dependence of the gain on the
densities in the source QW and top QW is more pronounced. In particular,
absorption occurs already for top-QW carrier densities below $10^{10}$~cm$^{-2}$, whereas for structure A there is still gain in this top-QW density range.

\subsection{Inversion (gain) with carrier injection\label{numerical2}}     

The above numerical results show that a carrier population in the top QW, i.e., the QD scattering states, has a detrimental effect on the gain. Further, it is shown in the present section that an accumulation of carriers in the top QW  precludes a steady-state inversion (gain) in structures A and B, if one includes a model for carrier injection. To reach a steady-state gain one therefore has to counteract the piling up of population in the top QW. We propose to achieve this by removing carriers from these states directly as described in Sec.~\ref{gainmaterial1}, and analyze the dynamics with the additional carrier extraction in some detail. We will do  these calculations for structure B because in that structure source and top QW states are separated by a barrier so that source and top QW can be better addressed separately by an injection/extraction process. For comparison we will also analyze the behavior of structure A with carrier injection, but we will always assume only extraction from the drain QW for structure A.

\begin{figure}[tb]
\centering
\includegraphics[scale=0.4]{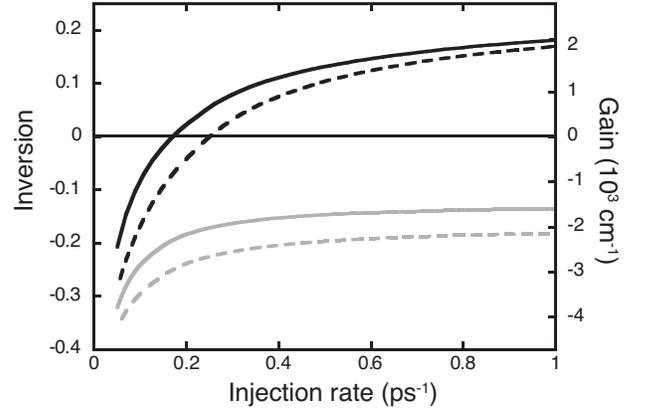}
\caption{Population inversion and gain for the QD transitions versus
injection rate for structure~A (gray lines) and structure~B (black lines). The lattice temperature is 150\,~K (solid
line) and 300\,K (dashed line).}
\label{num2_fig1and2}
\end{figure}

The basic dynamical equations are Eqs.~\eqref{electron-phonon}
and~\eqref{electron-electron} for electron-phonon and electron-electron
scattering, but now including carrier injection terms~\eqref{midlas_inject}
and \eqref{midlas_extract}. In particular, the processes (i), (iii) and (iv) depicted in
Fig.~\ref{skizze_QWQDQW} are now considered. The pump distribution $F_{k}$ of the
injection (extraction) process depends on the particular device in which the
QW-QD-QW structure is embedded. Unless otherwise specified, we assume
$N^{F}=5\times 10^{10}$~$\text{cm}^{-2}$ and lattice temperatures of
$T^{F}=300$\,K and $T^{F}=150$\,K, respectively. Note that in addition to intra-QD electron scattering processes and QD-QW electron scattering processes, also intra-QW electron scattering
processes occur. All the following numerical results are again computed starting from an initially empty QW and QD system and evolve the 
carrier distribution until a steady-state is reached. 

We first investigate whether a steady-state inversion, i.e., gain, can be achieved for structure A or B.
Figure~\ref{num2_fig1and2} plots the population inversion and gain for the
QD transition versus the injection rate for structure A and B. For structure A the inversion
rises with increasing injection rates but saturates at negative values for a lattice temperatures of
$300$\,K and for $150$\,K. The inversion for a lattice temperature of $150$\,K exceeds that for $300$\,K at all injection rates. This can be expected from the increased efficiency of electron-phonon relaxation between the QD states at higher temperatures, which works against an inversion on the QD intersubband transition. However,  the difference becomes smaller with increasing injection rate.
For structure B the inversion also rises with increasing injection rate and reach a saturation value, which is \emph{positive}: For injection rates above 
1.2\,ps$^{-1}$ a saturation value of the inversion around $0.2$ is reached.

\begin{figure}[tb]
\centering
\includegraphics[scale=0.4]{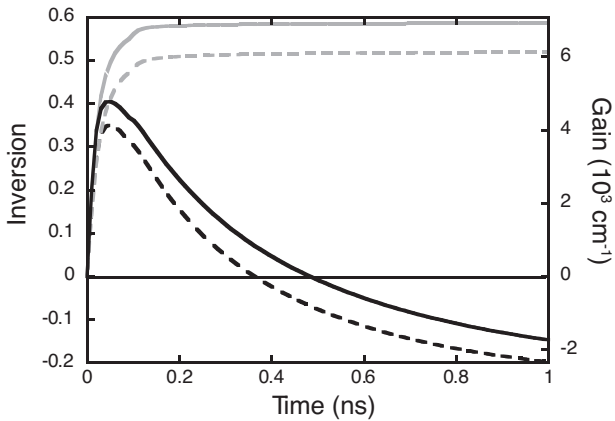}
\caption{Population inversion and gain for the QD transition versus
time for structure~A. The full calculation including electron-electron and electron phonon scattering (black lines) is compared with the result including only electron-phonon scattering (gray lines). The lattice temperature is 150\,K (solid lines) and 300\,K (dashed lines).}
\label{num2_fig3}
\end{figure}

\begin{figure}[tb]
\centering
\includegraphics[scale=0.4]{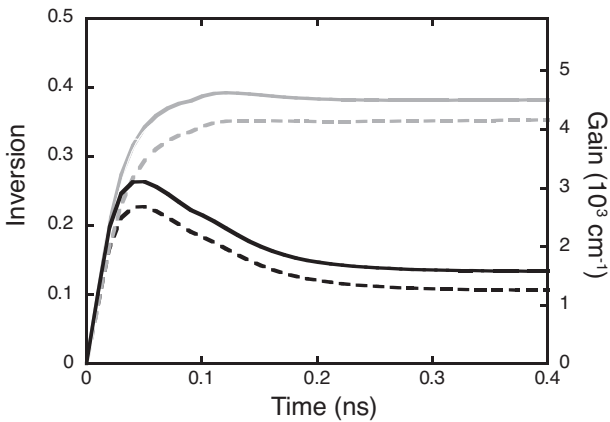}
\caption{Same as Fig.~\ref{num2_fig3} for structure~B.}
\label{num2_fig4}
\end{figure}

For a more detailed analysis of these results, in Fig.~\ref{num2_fig3} and Fig.~\ref{num2_fig4} we look at the time dependence of the population inversion for a fixed injection rate of $\Lambda=0.5\,\text{ps}^{-1}$ for structure A and B, respectively. A calculation including both electron-phonon and electron-electron scattering (``ep+ee'') is compared to a calculation including only electron-phonon scattering (``ep''). Both calculations are done for lattice temperatures of 150\,K and 300\,K. 

Fig.~\ref{num2_fig3} plots the population inversion for the QD transition
versus time for structure A. As long as the top-QW states are essentially
empty, the ep+ee and the ep results are very similar. After a few tens of
picoseconds the top-QW states are significantly populated, and
electron-electron scattering becomes more efficient for the
dynamics. As already discussed in Sec.~\ref{numerical1} top-QW assisted QD
electron relaxation becomes more important. Further, source-QW assisted QD electron capture and source-QW assisted QD
electron relaxation contribute to different results for the inversion. In addition, the electron-electron scattering leads to a faster and more homogeneous redistribution of carriers in the QWs. Taken together, very different electronic distributions (with different electron densities) are reached after a few ns. The net effect is that the achievable inversion $N$ is negative for the ep+ee and positive for the ep calculation in steady-state.

Figure~\ref{num2_fig4} shows the same plot for structure B. The ep
calculation for structure B is similar to that of structure A shown in
Fig.~\ref{num2_fig3}, with structure A leading to higher gain (inversion). The
important difference is between the ``full'', namely, ep+ee, calculations. Here,
the initial dynamics over a few tens of ps is similar to that of structure A,
but much different when the carrier density rises and the influence of
electron-electron scattering becomes pronounced. Since the extraction from
drain \emph{and top QW} states limits the carrier density in the drain
\emph{and top QW}, the  inversion $N$ remains positive for all times and
leads to a positive gain in steady state. As already mentioned in
Sec.~\ref{numerical1} above, structure A performs better for fixed carrier
densities in the source QW. But if a carrier injection model is included, only
in structure B (with carrier extraction from the top-QW states) steady-state gain can be realized.  We will therefore focus on structure B in the following.

\begin{figure}[tb]
\centering
\includegraphics[scale=0.4]{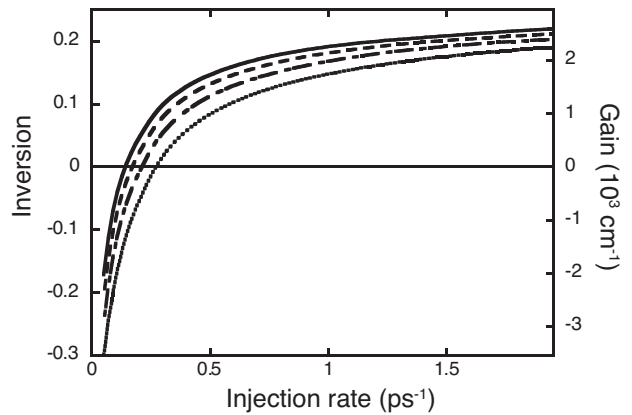}
\caption{Population inversion and gain for the QD transitions versus
injection rate for structure B. The carrier density refering the pump
distribution is $N^{F}=6\times10^{10}$~$\text{cm}^{-2}$ (solid line),
$N^{F}=5\times10^{10}$~$\text{cm}^{-2}$ (dashes line),
$N^{F}=4\times10^{10}$~$\text{cm}^{-2}$ (dot-dashed line) and
$N^{F}=3\times10^{10}$~$\text{cm}^{-2}$ (dotted line). The lattice temperature is 150\,K.}
\label{num2_fig5}
\end{figure}

We also investigate how the carrier density of the pump distribution $N^{F}$ affects the 
results. As already mentioned, we treat the pump distribution as a parameter. Fig.~\ref{num2_fig5} shows the population inversion and gain for the QD transitions versus injection rate for structure B for different carrier densities
$N^{F}$ and a lattice temperature of 150\,K. More precisely, we choose $N^{F}=6\times10^{10}$~$\text{cm}^{-2}$,
$N^{F}=5\times10^{10}$~$\text{cm}^{-2}$, $N^{F}=4\times10^{10}$~$\text{cm}^{-2}$ and
$N^{F}=3\times10^{10}$~$\text{cm}^{-2}$ for a comparison.
For larger $N^{F}$, lower injection rates are necessary to achieve similar
gain values. However, apart from that, the $N^F$ has no decisive influence on the gain ``dynamics''. Thus the variation of injection rate allows one to determine the important characteristics of the QW-QD-QW active region.

\subsection{Strong-signal effects\label{numerical3}}

In this section we go beyond small-signal gain results by including an
externally controlled optical field. The optical field may be the field in an optical amplifier or
in a laser cavity. We run a dynamical calculation for the densities
and the optical polarizations based on the semiconductor Bloch equations,
i.e., \eqref{p-aa} and \eqref{n-a}. Again electron-phonon and
electron-electron scattering is included for the whole system under
investigation, in particular, the processes (i)-(iv) depicted in
Fig.~\ref{skizze_QWQDQW} contribute. We are interested in the dependence of the steady-state
inversion $N$, or equivalently the gain $G$, see Eq.~\eqref{midlas_gain}, on the optical field intensity, and we analyze exclusively structure B.
 
\begin{figure}[tb]
\centering
\includegraphics[scale=0.4]{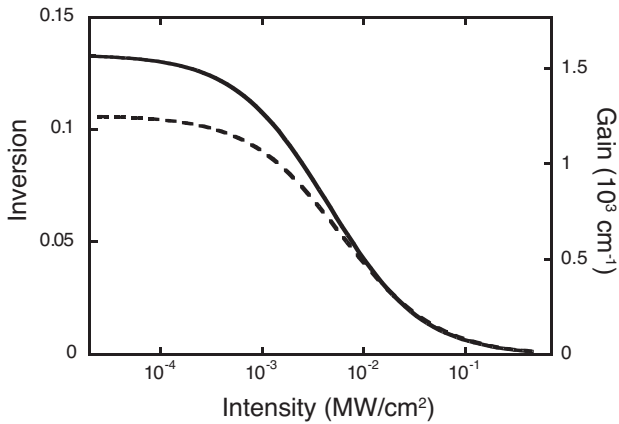}
\caption{Population inversion and gain for the QD states versus
field intensity for the structure B. Lattice temperatures
are 150~K (solid line) and 300~K (dashed line).}
\label{num3_fig1}
\end{figure}

The inversion and gain achievable with structure B versus field intensity for a lattice temperature
of $300$\,K and $150$\,K are depicted in Fig.~\ref{num3_fig1}. We assumed
a fixed injection rate of $0.5$ $\text{ps}^{-1}$ with $N^{F}=5\times 10^{10}$~$\text{cm}^{-2}$ for
the pump distribution. The weak field result is
recovered for small field intensities below $10^{-4}$~$\text{MW/cm}^{2}$, 
as it should be. For increasing field intensity the
inversion and the gain decrease, because the optical field leads to a
stimulated recombination of carriers and reduces the inversion. For field intensities
between $10^{-4}$~$\text{MW/cm}^{2}$ and $10^{-1}$~$\text{MW/cm}^{2}$ the
inversion, i.e. gain, is still positive,
but decreases rapidly. For field intensities above
$10^{-1}$~$\text{MW/cm}^{2}$ no significant inversion or gain is observed.
While for weak field intensities the lower lattice temperatures has the higher gain, this difference is strongly reduced with increasing intensity. Above $10^{-2}$\,MW/cm$^{-2}$ the gain curves for the two temperatures are almost indistinguishable, with the gain in the high temperature case being even slightly higher. This can be explained as follows: For small field intensities a lower lattice temperature leads to a
higher carrier density at the band edge of the source QW, and thus to a higher
steady-state population of the excited QD states and consequently a higher
inversion. For higher field intensities, the scattering between
the band edge of the source QW and the excited QD states needs to be more
efficient to sustain to the same inversion, so that now the scattering efficiency also plays a more important role, in addition to the population of the QW states. The scattering efficiency is
higher for higher lattice temperatures, because electron-phonon scattering is
more efficient due to polaronic state broadening effects. This leads to very similar 
gain for higher field intensities for different lattice temperatures.

These results with a fixed optical field intensity can be used as a figure of
merit for the performance of the QW-QD-QW structure as a laser gain material:
If the cavity losses of a particular laser structure are known, this
determines the saturated gain in steady-state. 
The extracted values are for the saturated gain, i.e., the gain of the
active region and not the modal gain for a specific device, see Sec.~\ref{numerical2add2}. 
However, from the results of Fig.~\ref{num3_fig1} an estimate of the intensity of the
optical field in the cavity is possible, for instance, for an injection rate of $0.5\,\text{ps}^{-1}$.

\subsection{Dependence on nonuniform injection (extraction)
  rates\label{numerical2add}}

So far we have assumed equal injection (extraction) rates for all three QWs. In this
section we investigate the dependence of the gain for nonuniform injection (extraction)
rates for structure B. Therefore we distinguish between the injection into the source QW,
$\Lambda_{s}$, the QW extraction from the drain QW, $\Lambda_{d}$, and the
extraction from the top QW, $\Lambda_{t}$. As analyzed in Sec.~\ref{numerical2}, with increasing injection the gain reaches a positive saturation value. At the onset of saturation, i.e. for injection (extraction) rates of $\Lambda_{s}=\Lambda_{d}=\Lambda_{t}=1.0$\,ps$^{-1}$, the
inversion is approximately 0.18 as shown in Fig.~\ref{num2_fig1and2}. In the
following we vary the different injection (extraction) rates around this configuration.

\begin{figure}[tb]
\centering
\includegraphics[trim=4cm 1cm 4cm 2cm,clip,scale=0.4]{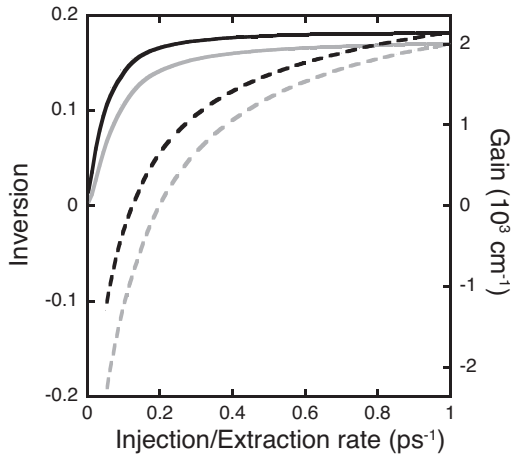}
\caption{Population inversion (and gain) for the QD transitions versus
injection rate into the source and equal extraction rate from the drain QW
respectively (solid line) and
extraction rate from the top QW (dashed line). The lattice temperature is
150\,~K (black) and 300\,~K (gray).}
\label{num2b_fig4}
\end{figure}

We start by changing $\Lambda_{s}$ and $\Lambda_{d}$ together and keep
$\Lambda_{t}=1.0$\,ps$^{-1}$ constant.
The numerical calculation is done as already described in Sec.~\ref{numerical2} and the carrier distributions are evolved until a
steady-state is reached. The result is shown in
Fig.~\ref{num2b_fig4} for a lattice temperature of $150$\,K and $300$\,K. The
inversion curve rises with increasing injection (extraction)
rates and goes into saturation around $0.1$\,ps$^{-1}$, which is
below the values found for equal rates. Thus, it is possible to reduce the injection (extraction) rate
into the source and drain QW, if the extraction rate of the top QW is kept
constant at 1\,ps$^{-1}$. In particular,  we obtain positive gain
for all injection (extraction) rates.

In the next step we vary $\Lambda_{t}$ and keep
$\Lambda_{s}=\Lambda_{d}=1.0$\,ps$^{-1}$ constant. The result is also depicted
in Fig.~\ref{num2b_fig4} for a lattice temperature of 150\,K and
300\,K. Gain saturation is reached around $1.0$\,ps$^{-1}$, which was already
found in Sec.~\ref{numerical2}. In particular, transparency is reached at similar rates
in the two cases. This suggests that the qualitative behavior in figure
\ref{num2_fig1and2} is dominated by the extraction rate of the top QW. 

\begin{figure}[tb]
\centering
\includegraphics[scale=0.4]{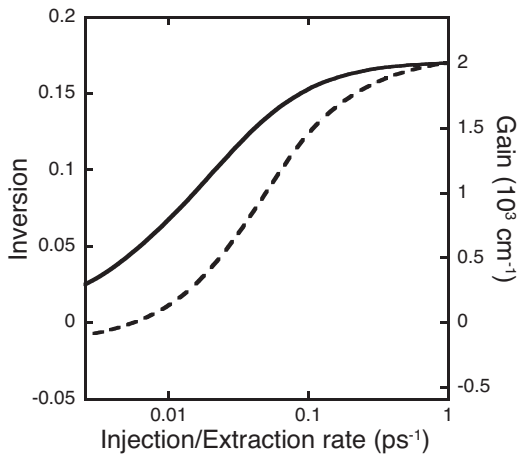}
\caption{Population inversion (and gain) for the QD transitions versus
injection rate into the source QW (solid line) and extraction rate from the drain
QW (dashed line). The lattice temperature is 300\,~K.}
\label{num2b_fig5}
\end{figure}

Finally we vary the ratio between $\Lambda_{s}$ and  $\Lambda_{d}$, while keeping
$\Lambda_{t}$ constant. The results are depicted in Fig.~\ref{num2b_fig5}
for a lattice temperature of 300\,K. For 150\,K the results are qualitatively
similar. If only $\Lambda_{s}$ is varied, the saturation is reached
around $0.1$\,ps$^{-1}$ and the gain remains positive for all
injection rates. If only $\Lambda_{d}$ is changed, the inversion is
more sensitive to this change (note the logarithmic plot), but still reaches comparable values already
around an extraction rate of $0.1$\,ps$^{-1}$. Only for a constant
$\Lambda_{s}=1.0$\,ps$^{-1}$ and a very low extraction rate the gain can be negative,
because the carriers are not extracted sufficiently fast from the drain QW. Note that the
gain remains positive for all other combinations of these two rates. To sum up, the dependence of the gain on $\Lambda_{s}$
and $\Lambda_{d}$ is similar, and starting from an equal injection (extraction)
rate $\Lambda=1.0$\,ps$^{-1}$ the gain is robust against a reduction of $\Lambda_{s}$
or $\Lambda_{d}$. 

For a QCL design it might be important to know the ratio between top and drain QW
carrier extraction. A calculation of this ratio in the framework of our model shows that 
leakage through top QW extraction generally stays around five percent. To avoid a reduction of differential quantum efficiency in a
QD based QCL the collector region of the QCL (see section \ref{numerical2add2}) should support
the relaxation of the extracted top QW carriers into the following source QW.

\section{Comparison between a QD- and QW-QCL \label{numerical2add2}}

In our investigation of AlGaAs/InGaAs QW-QD-QW structures as active material
for midinfrared lasers, the design of the surrounding
heterostructures (e.g. collector region) has not been taken into account, viz., we do not investigate a device model of a
QD-QCL. However, the results of the analysis done are transferable to a
periodic structure (like a QD-QCL). For that purpose a collector region
between the drain QW and the source QW has
to be added. In this region the carriers from the top and drain QW are collected and
injected into the subsequent source QW. We assume that all carriers extracted
from the precedent top and drain QWs are injected into the subsequent
source QW, i.e. carrier losses in the collector region are neglected.
Under this assumption the steady-state current density~$J$ through the QD-QCL device can
be determined as described in Sec.~\ref{subsec:carrier-injection}. In the following we compare our results to those of a QW-QCL investigated
in Ref.~\onlinecite{Harrison1}. Therefore we choose an analogous confinement factor of $\Gamma_{\text{con}}=0.42$ and a similar device
periodicity of $50$~nm, which corresponds to a field around 36\,kV/cm for our
structure. The small-signal modal gain $G_{M}$ is given by 
\begin{equation}
G_{M}=\Gamma_{\text{con}}2\frac{\omega }{c_{0}\varepsilon _{0}n_{b}}\frac{\Gamma_{\text{inh}} N_{D}}{L_{\text{P}}} \frac{\mu ^{2}}{\hbar \gamma _{d}}N  
\label{modal_gain}
\end{equation}
where $L_{\text{P}} = 50$~nm is the periodicity length of the structure. The other parameters are the same as in Sec.~\ref{numerical1}, see
Eq.~\eqref{midlas_gain}). Here, we have also included an inhomogeneous broadening $\Gamma_{\text{inh}}$ of the QD ensemble. While the polarization dephasing determines the homogeneous broadening, 
the inhomogeneous broadening acts as an effective reduction of the density $N_{D}$ of QDs that are resonant with the optical field. 

\begin{figure}[tb]
\centering
\includegraphics[scale=0.4]{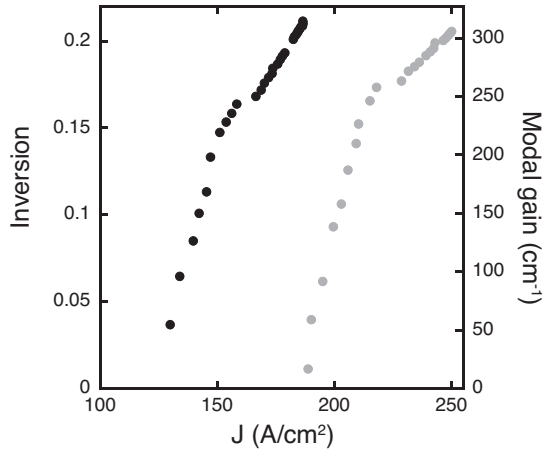}
\caption{Modal gain versus current density for 150~K (black dots)
  and 300~K (gray dots) without inhomogeneous broadening.}
\label{num2c_fig1}
\end{figure}

The carrier distributions for variable uniform injection rates are evolved
until a steady-state is reached and the steady-state injection current density
into the source QW and the steady-state modal gain (calculated from
the inversion $N$) is determined. In Fig.~\ref{num2c_fig1} the modal gain
versus current density for different injection rates is plotted for a lattice
temperature of 150~K and 300~K without inhomogeneous broadening. 
Qualitatively, a higher current density leads to a higher modal gain. For a
lattice temperature of 300~K, a higher current density is needed to obtain the
same modal gain. In particular, for a lattice temperature of 150~K and an injection rate of $\Lambda = 1.0$\,ps$^{-1}$, a steady-state current density
of $J = 180$\,A~cm$^{-2}$ and an inversion close to saturation of N$= 0.17$
is reached. For a lattice temperature of 300~K the same inversion is
reached for an injection rate of $\Lambda = 1$\,ps$^{-1}$ and a steady-state current density
of $J = 220$\,A~cm$^{-2}$.

\begin{figure}[tb]
\centering
\includegraphics[scale=0.4]{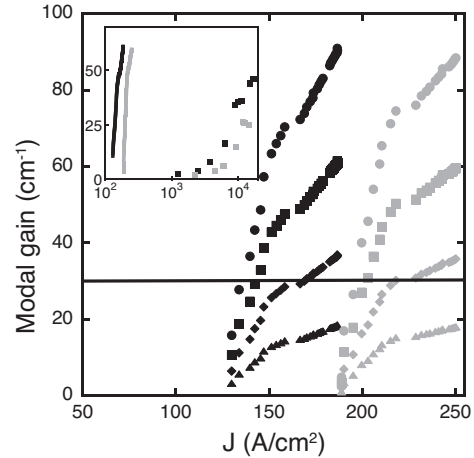}
\caption{Modal gain versus current density for 150~K (black)
  and 300~K (gray). The FWHM of the inhomogeneous broadening is 10~meV
  (dots), 15~meV (rectangles), 25~meV (diamonds), 50~meV (triangles). The
  black line is the total loss line. Inset: Comparison of modal gain versus
  current density between the QD model of this paper and a QW-QCL from
  Ref.~\onlinecite{Harrison1}. For the QD model, we replot the data for 150~K
  (black line) and 300~K (gray line) with a broadening of 15~meV. For the QW-QCL the data for 200~K (black
  rectangles) and 300~K (gray rectangles) are taken from Fig.~14 of Ref.~\onlinecite{Harrison1}.}
\label{num2c_fig2}
\end{figure}

In Fig.~\ref{num2c_fig2} the modal gain
versus current density for different injection rates is plotted for lattice
temperatures of 150~K and 300~K. The inhomogeneous broadening is included via a Gaussian
profile with FWHM of 10~meV, 15~meV, 25~meV and 50~meV. The total loss line $\alpha$ is chosen in agreement with Ref.~\onlinecite{Harrison1} as $\alpha = 30$~cm$^{-1}$. It is a summation of the mirror and waveguide
losses. An inhomogeneous broadening with a FWHM smaller than 25~meV is necessary to overcome
the total losses $\alpha$. For an inhomogeneous broadening with FWHM between
10~meV and 25~meV and a lattice temperature of 150~K a threshold current
density around $J = 140$\,A~cm$^{-2}$ and for a lattice temperature of 300~K a threshold current
density around $J = 200$\,A~cm$^{-2}$ is needed. 
In comparison to the results obtained for a QW-QCL structure investigated in
Ref.~\onlinecite{Harrison1} (their Fig.~14), which are displayed in the inset of
Fig.~\ref{num2c_fig2}, the threshold current density is approximately 50 times lower in our QD-QCL
structure. However, a comparable gain can be only achieved for an inhomogeneous
broadening of the QD ensemble that is close to what is achievable by self-organized growth at present.
More recent experimental results for QW-QCL devices reach threshold current
densities of about $1$\,kA~cm$^{-2}$. In particular, in Ref.~\onlinecite{Slivken} a
threshold current density as low as $810\,\text{A}\text{cm}^{-2}$ has been measured for a device with a
smaller total loss and a larger confinement factor than assumed in our calculation. A direct comparison to these more recent experimental devices
leads to threshold reduction of roughly a factor of 5. 

\section{Conclusion}

In conclusion, we presented a microscopic calculation for the
gain arising from intersubband transitions in QDs in the mid-infrared
range. In order to provide a realistic description of how inversion on an
electronic intersubband transition in QDs can be achieved, we assumed that a
QD layer was sandwiched between a source and a drain QW, and we modeled the
carrier injection and extraction into the QWs, respectively. We included a
realistic description of the QD electronic structure and a microscopic treatment
of electron-phonon and electron-electron scattering. 
We analyzed two structures, which differed mainly in a separation of the
source QW from the QD and top QW. It was found that substantial gain can
only be achieved if one allows for direct carrier extraction from the
scattering continuum of the QDs, which is only possible if the source QW is
separated from the QD and drain as well as the \emph{top QW}. Only in this case the
scattering states above the QD do not become substantially occupied by the
injection. If the population of the scattering states is too large, these
electrons act as scattering partners for electrons in the localized QD states,
and lead to a more efficient relaxation towards the QD ground state, thus
decreasing the inversion in the QD. For the optimized structure significant gain is found in the small signal limit 
as well as beyond the small signal limit up to $0.1\,\text{MW/cm}^{2}$.
For higher field intensities the gain of the QD intersubband transition is depleted. The
dependence of the gain versus field intensity can be used as a figure of merit for the performance as gain material in a laser.
In addition, the tradeoff between the different injection (extraction)
pathways was analyzed and potential leakage pathways were discussed. We found that the rates are dominated by the extraction rate of the
top QW and the ratio between top and drain QW carrier extraction is around five percent. 
Finally, we compared our QD-QCL to a standard QW-QCL device as analyzed in
Ref.~\onlinecite{Harrison1} and more recent experimental results.~\cite{Slivken} The threshold current densities predicted for the QD-QCL structure are reduced in comparison to QW-based designs, but a comparable modal gain for the QW- and QD-QCL structure is possible only for an inhomogeneous broadening of the QD ensemble that is close to what is achievable today.

\begin{acknowledgments}
This work was supported in part by Sandia's Solid-State Lighting Science
Center, an Energy Frontier Research Center (EFRC) funded by the US Department
of Energy, Office of Science, Office of Basic Energy Sciences. 
\end{acknowledgments}

\appendix 

\section{Calculation of the electronic structure and the Coulomb- or
  carrier-phonon-scattering matrix-elements\label{a_mparameter}}

The electronic structure consisting of conduction-band QW and QD states is
calculated by k$\cdot$p theory.
We calculated the one-dimensional envelopes of the QWs
$\xi^{b_c} \left( z \right) $ and
the three-dimensional QD states $\Psi^{b_c}_m \left( x,y,z \right) $ 
in a single-band approximation using the
software package in Ref. \onlinecite{homepage}.
The values for material parameters of AlGaAs and InGaAs compounds are taken from Ref. \onlinecite{material}.
This approach cannot handle the whole system in
one ``box,'' which would yield localized and delocalized eigenfunctions that are orthogonal
to each other.
Instead, we extend the one-dimensional envelopes of the QWs to three-dimensional QW states
\begin{equation}
\Psi^{b_c}_{k,\phi} \left( x,y,z \right) = \xi^{b_c} \left( z \right) \xi_{k,\phi} \left( x,y \right) 
\end{equation} 
assuming a parabolic conduction band with plane waves $\xi_{k,\phi} \left( x,y \right)$ as in-plane
functions and $\phi$-independent energy values for the QW states $\Psi^{b_c}_{k,\phi}$. 
To describe the combined system we orthogonalize the QW states to the QD states with
\begin{equation}
\begin{split}
\Psi^{\perp,b_c}_{k,\phi} \left( \vec{r} \right) =& \frac{1}{N_k} \Psi^{b_c}_{k,\phi} ( \vec{r}) \\
	& - \frac{1}{N_k} \sum_m  \Psi^{b_c}_m (\vec{r}) \int d^{3}{r'} 
		\Psi^{b_c \ast}_m (\vec{r}')  \Psi^{b_c}_{k,\phi} (\vec{r}')
\end{split}
\end{equation}
where $N_{\vec{k}}$ is a normalization constant.
The outcome of this are localized and delocalized eigenfunctions that are orthogonal
to each other.~\cite{ortho}

For the following explanations its useful to simplify the notation of the
band index.
Here, we investigate a QD d of the ensemble with $M^{e}$ electron states
embedded in a QW structure consisting of a source-QW S, a
top-QW W and a drain-QW D. Especially, in a single-band approximation
for the conduction band c where all electron states are spin degenerate, every state in the QD
can be labeled by $\lambda =( b,\vec{k}=m,s) $ where $b =
( \text{d,c}) $ is a generalized band index, $m\in \{ 1,\ldots
,M^{e}\}$ is a QD state index and $s\in \{ \uparrow
,\downarrow\} $ is the spin index. States in the QWs are labeled by $\lambda
=( b,\vec{k}=\vec{k}_{\parallel },s) $ where $b\in
\{ ( \text{S,c} ) , ( \text{W,c}) , ( \text{D,c} ) \} $ is a generalized band index for
the source-, top- and drain-QW. Thus we introduce
the notation $\lambda _{1}$ with $\lambda _{1}=( b_{1},\vec{k}_{1},s_{1}) $ for all states. With this unified index $\lambda _{1}=( b_{1},\vec{k}_{1},s_{1}) $ a simplified notation of the
carrier-phonon interaction matrix-elements $M_{\lambda _{2},\lambda _{1}}$ 
and the carrier-carrier interaction
matrix-elements $W_{\lambda _{3}\lambda_{4}}^{\lambda _{1}\lambda _{2}}$ follows. 

The electron-electron and electron-phonon scattering contributions are gathered in
appendix \ref{a_scatter}. Here, we are concerned with the computation of $M_{\lambda _{2},\lambda _{1}}$ 
and $W_{\lambda _{3}\lambda _{4}}^{\lambda _{1}\lambda _{2}}$.
The carrier-carrier interaction matrix-elements are calculated using
\begin{equation}
W^{\lambda_1,\lambda_3}_{\lambda_2, \lambda_4} = \frac{1}{V} \sum_{\vec{q}} w_{\vec{q}}  I^{\lambda_1}_{\lambda_2}( \vec{q}) I^{\lambda_3}_{\lambda_4}(-\vec{q})
\label{W_elel}
\end{equation}
where 
\begin{eqnarray}
I^{\lambda_1}_{\lambda_2}\left( \vec{q} \right) = \int d^3\vec{r} \text{ }
\Psi^{\star}_{\lambda_1} \left( \vec{r} \right) e^{i\vec{q}\vec{r}}
\text{ }\Psi_{\lambda_2} \left( \vec{r} \right) \\
I^{\lambda_1}_{\lambda_2}\left( -\vec{q} \right) = \int d^3\vec{r} \text{ }
\Psi^{\star}_{\lambda_1} \left( \vec{r} \right) e^{-i\vec{q}\vec{r}}
\text{ }\Psi_{\lambda_2} \left( \vec{r} \right)
\end{eqnarray}
In the numerical implementation of the electron-electron scattering, the
Coulomb-matrix elements $W^{\lambda_1,\lambda_3}_{\lambda_2,\lambda_4}$
including the integrals $I^{\lambda_1}_{\lambda_2}$ are part of an
integral-kernel expression $I_{\text{el}} (k_1, k_2, k_3, k_4) $,
which is independent of the angle $\phi$ of $\vec{k}_{\parallel}$. For the calculation of $I_{\text{el}}$, $\vec{q}$ has cylindrical
coordinates, because they are well suited for the evaluation of our QW system
with embedded QD states.

For QD-QD, QW-QD and QD-QW integrals $I^{\lambda_1}_{\lambda_2}(\vec{q})$ are calculated numerically because the wave-function overlaps are
finite in all three dimensions.    
QW-QW integrals $I^{\lambda_1}_{\lambda_2}\left( \vec{q} \right)$ has to be
calculated semi-analytically similar to Ref.~\onlinecite{Jahnke1}, because the integral
components related to the in-plane functions resulting in $\delta$-functions
which has to be included into $W^{\lambda_1,\lambda_3}_{\lambda_2,\lambda_4}$
or $I_{\text{el}}$ as analytical expressions.   
The Coulombmatrix $W^{\lambda_1,\lambda_3}_{\lambda_2,\lambda_4}$ has to be
interpreted by an distinction between different combinations of QW and QD states. 
More precisely, we distinguish between
intra-QW scattering (4 QW states), QW-assisted QD capture/emission (3 QW states), 
QW-assisted QD scattering (2 QW states paired), pure QD-QW scattering (2 QW states unpaired),
QD-assisted QD capture/emission (3 QD states) and intra-QD scattering (4 QD states).
Finally, the Coulombmatrix $W^{\lambda_1,\lambda_3}_{\lambda_2,\lambda_4}$ is
included into the electron-electron scattering integral-kernel $I_{\text{el}}$
where all integrals over the $\phi$'s, i.e. $\vec{k}_{\parallel}$ angles, are evaluated. 

In the numerical implementation of the electron-phonon scattering, the
carrier-phonon interaction matrix-elements $M_{\lambda _{2},\lambda _{1}}$ 
are also part of an integral-kernel expression $I_{\text{ph}} \left( k_1, k_2 \right) $,
which is independent of the angle $\phi$ of $\vec{k}_{\parallel}$. 
Especially, the electron-phonon scattering integral-kernel $I_\text{ph}$
contains the expression 
\begin{equation}
\sum_{\vec{q}}\left\vert M_{\lambda _{2},\lambda _{1}}(\vec{q}) \right\vert ^{2}
= \frac{1}{V} \sum_{\vec{q}}
\frac{M_{\text{LO}}^{2}}{\vec{q}^2}  I^{\lambda_2}_{\lambda_1}\left( \vec{q} \right)
I^{\lambda_1}_{\lambda_2}\left( -\vec{q} \right)
\label{M_elph}
\end{equation}
where $M_{\text{LO}}$ is the prefactor of the Froehlich Hamiltonian.
We evaluated $I_\text{ph}$ analog to $I_\text{el}$, because the expressions
in $I_\text{ph}$ resulting from Equ. (\ref{M_elph}) can be treated comparable to the
expressions in $I_\text{el}$ resulting from Eq.~\eqref{W_elel}. 
For the carrier-phonon interaction matrix-elements $M_{\lambda _{2},\lambda _{1}}$
we can distinguish between
intra-QW scattering (2 QW states), phonon-assisted QD capture/emission (1 QW
state) and intra-QD scattering (2 QD states).

\section{Scattering contributions\label{a_scatter}}

We calculate the electron densities $n_{\lambda _{1}}$
for the whole system under investigation dynamically. Thus,
electron-phonon and electron-electron scattering terms 
including both QW- and QD-states.
For the analysis of the scattering contributions 
we distinguish between intra-QW electron scattering, intra-QD electron scattering
and QD-QW scattering processes, but summarize the explicit expressions with an
unified index.
More precisely, we refer to
intra-QW electron-electron and electron-phonon scattering as intra-QW
electron scattering processes and to
intra-QD electron-electron and electron-phonon scattering as intra-QD
electron scattering processes.
Further, we summarize
QW-, QD-, or phonon-assisted QD electron capture/emission, QW-assisted
QD-scattering and pure QD-QW scattering and refer to them as QD-QW electron scattering processes.
However, for the explicit scattering contributions we use our unified index $\lambda _{1}=( b_{1},
\vec{k}_{1},s_{1}) $ as introduced in appendix~\ref{a_mparameter}.
A simplified notation of the carrier-phonon interaction matrix-elements $M_{\lambda _{2},\lambda _{1}}$ and the carrier-carrier interaction
matrix-elements $W_{\lambda _{3}\lambda _{4}}^{\lambda _{1}\lambda _{2}}$ follows.

\begin{widetext}

The derivation of the scattering contributions is described in Ref.~\onlinecite{QDM}.
In contrast to Ref.~\onlinecite{QDM} all coherences
are neglected and we assume that only conduction band states are involved in
the scattering process. With a generalized notation $n_{\lambda _{1}}$ for the electron densities
we obtain for scattering processes due to the carrier-phonon interaction
in Markov approximation

\begin{eqnarray}
S_{\lambda _{1}}^{cp} = \frac{2\pi }{\hbar }\sum_{\lambda _{2}}\left(
\sum_{\vec{q}}\left\vert M_{\lambda _{2},\lambda _{1}}\left( \vec{q}
\right) \right\vert ^{2}\right)
\left\{
\begin{array}{c}
\left[ 1-n_{\lambda _{1}}\left( t \right) \right] n_{\lambda
_{2}}\left( t \right)
\left(1+N\right)\widehat{g}\left( -\widetilde{\varepsilon }
_{\lambda _{2}}+\widetilde{\varepsilon }_{\lambda _{1}}^{\ast }+\hbar \omega
_{LO}\right) \\ 
+ \left[ 1-n_{\lambda _{1}}\left( t \right) \right] n_{\lambda
_{2}}\left( t \right)
\text{ }N\text{ }\widehat{g}\left( -\widetilde{\varepsilon }_{\lambda
_{2}}+\widetilde{\varepsilon }_{\lambda _{1}}^{\ast }-\hbar \omega
_{LO}\right)
 \\  
-n_{\lambda _{1}}\left( t \right)\left[ 1-n_{\lambda _{2}}\left(
t \right) \right]
N\text{ }\widehat{g}\left( -\widetilde{\varepsilon }_{\lambda _{2}}+
\widetilde{\varepsilon }_{\lambda _{1}}^{\ast }+\hbar \omega _{LO}\right) \\ 
-n_{\lambda _{1}}\left( t \right)\left[ 1-n_{\lambda _{2}}\left(
t \right) \right]
\left(1+N\right)\widehat{g}\left( -\widetilde{
\varepsilon }_{\lambda _{2}}+\widetilde{\varepsilon }_{\lambda _{1}}^{\ast
}-\hbar \omega _{LO}\right)
\end{array}
\right\} \label{electron-phonon}
\end{eqnarray}
where $\hat g$ denotes the real part $g\left( z \right) =\frac{i}{\pi z}$ and
$\widetilde{\varepsilon }_{\lambda _{1}}=\varepsilon _{\lambda
_{1}}+\Delta \varepsilon -i\Gamma $ can be understood as a complex
single-particle energy with an energy shift $\Delta \varepsilon$ and a damping $\Gamma $, 
i.e., broadening in energy, reflecting a
quasi-particle lifetime. This broadening is important for the discrete levels
of the QD and includes polaronic effects.

Further we evaluated an expression for scattering processes due
to carrier-carrier interaction as described in Ref. \onlinecite{QDM}.
In contrast to Ref. \onlinecite{QDM} again all coherences
are neglected and we assume that only conduction band states are involved in
the scattering process. We obtain for
the carrier-carrier interaction in Markov approximation

\begin{eqnarray}
S_{\lambda _{1}}^{cc} &=&\frac{2\pi }{\hbar }\sum_{\lambda _{2},\lambda _{3},\lambda
_{4}}W_{\lambda _{3}\lambda _{4}}^{\lambda _{1}\lambda _{2}}\left(
W_{\lambda _{3}\lambda _{4}}^{\ast ,\lambda _{1}\lambda _{2}}-W_{\lambda
_{3}\lambda _{2}}^{\ast ,\lambda _{1}\lambda _{4}}\right)
\left\{
\begin{array}{c}
n_{\lambda _{1}}\left( t\right) \left[ 1-n_{\lambda _{2}}\left(
t\right) \right] n_{\lambda _{3}}\left( t\right)\left[ 1-n_{\lambda
_{4}}\left( t\right) \right] \\ 
-\left[ 1-n_{\lambda _{1}}\left( t\right) \right] n_{\lambda
_{2}}\left( t\right) \left[ 1-n_{\lambda _{3}}\left( t\right) \right]
n_{\lambda _{4}}\left( t\right)%
\end{array}%
\right\} 
\widehat{g}\left( \widetilde{\varepsilon }_{\lambda
_{1}}^{\ast }-\widetilde{\varepsilon }_{\lambda _{2}}+\widetilde{\varepsilon 
}_{\lambda _{3}}^{\ast }-\widetilde{\varepsilon }_{\lambda _{4}}\right) 
\notag \\
\label{electron-electron}
\end{eqnarray}

\end{widetext}

\end{document}